\documentclass[
showpacs,
floatfix,
aps,prl,amsmath,
twocolumn,
superscriptaddress,
]{revtex4}

\usepackage{epsfig}
\usepackage{bm}
\usepackage{epsf}
\usepackage{array}

\usepackage[varg]{txfonts}

\newcommand{\be}{\begin{equation}}
\newcommand{\ee}{\end{equation}}

\newcommand{\bea}{\begin{eqnarray}}
\newcommand{\eea}{\end{eqnarray}}

\newcommand{\bes}{\begin{split}}
\newcommand{\ees}{\end{split}}

\newcommand{\req}[1]{Eq.~(\ref{#1})}
\newcommand{\reqs}[1]{Eqs.~(\ref{#1})}
\newcommand{\rref}[1]{(\ref{#1})}

\newcommand{\wc}{\omega_{\rm c}}

\newcommand{\Rc}{R_{\rm c}}
\newcommand{\vF}{v_{\rm F}}
\newcommand{\tautr}{\tau_{\rm tr}}
\newcommand{\tauin}{\tau_{\rm in}}
\newcommand{\tauint}{\tilde \tau_{\rm in}}
\newcommand{\tauq}{\tau_{0}}
\newcommand{\taul}{\tau_{\rm sm}}
\newcommand{\taus}{\tau_{\rm sh}}
\newcommand{\tautrl}{\tilde \tau_{\rm sm}}

\newcommand{\vare}{\varepsilon}

\begin{document}

\title{Non-linear Resistivity of a Two-Dimensional Electron Gas in a
  Magnetic Field}
\author{M. G. Vavilov}
\affiliation{Department of Physics, University of Wisconsin, Madison, WI 53706 }
\author{I. L.  Aleiner}
\affiliation{ Physics Department, Columbia University, New
  York, NY 10027 }
\author{L. I. Glazman}
\affiliation{Theoretical Physics Institute, University of Minnesota, Minneapolis, MN 55455 }

\date{November 5, 2006}

\begin{abstract}
We develop a theory of nonlinear response to an electric field of a
two-dimensional electron gas (2DEG) placed in a classically strong
magnetic field. The latter leads to a non-linear current-voltage
characteristic at a relatively weak electric field. The origin of
the non-linearity is two-fold: the formation of a non-equilibrium
electron distribution function, and the geometrical resonance in the
inter-Landau-levels transitions rates. We find the dependence of the
current-voltage characteristics on the electron relaxation rates
in the 2DEG.
\end{abstract}
\maketitle

A magnetic field applied to a two-dimensional electron gas (2DEG)
changes the energy spectrum and
dynamics of electrons. It leads to a modification of the transport
characteristics of the 2DEG even at relatively weak magnetic fields, at
which the Landau levels~\cite{Ando} are not resolved yet and
the Quantum Hall effect~\cite{prange} is not developed.
The most well-studied modification of that
kind is the Shubnikov -- de Haas (SdH) oscillations in the resistivity
of 2DEG. Its observation, however, is restricted to fairly low
temperatures, $T\lesssim\hbar\wc$, so that the thermal broadening
$T$ of the electron distribution is small compared to the Landau
quantization energy $\hbar\wc$. In the low-temperature limit, the
strength of the oscillations is controlled by the Dingle factor,
$\lambda= [-\pi/(\wc\tauq)]$; it yields information about the
``quantum'' lifetime $\tauq$ of the electron~\cite{Ando}
due to scattering off disorder.

Recently it was realized~\cite{du02} that the effect of a magnetic
field on the dc non-linear transport, unlike the SdH oscillations of
the linear resistivity, is not confined to low temperatures.
Oscillations of the differential resistivity with the magnetic field
at a finite level of current observed in Ref.~\cite{du02} persisted to
quite high temperatures (about $4$K), while the conventional SdH
effect was fully smeared out by temperature. This finding was
confirmed in later experiments~\cite{vitkalov05,vitkalov06,zudov06}
where the differential resistivity was measured both as a function of
the applied magnetic field and transport current. The oscillations of
the nonlinear resistance were associated with the geometrical
resonance in the electron transitions between the Landau
levels~\cite{du02} that arises from the commensurability of
the period in the spatial oscillations of the density of states (DOS) and
the diameter $2\Rc$ of an electron cyclotron trajectory.
Although this was a plausible explanation of the
effect, it remained unclear, why the oscillations are so weakly
sensitive to the temperature and which parameters of the 2DEG control
the amplitude of the oscillations.

Another notable effect of magnetic fields on the
nonlinear transport in 2DEG was reported in Ref.~\cite{vitkalov06} and
deals with the region of relatively small current densities. In that
regime, a sharp drop in the differential resistivity was observed.
The effect was attributed to the modification of the electron energy
distribution caused by the current~\cite{DVAMP}. Clearly, this modification
depends on the energy relaxation rate, and the mechanisms behind the
observations reported in Refs.~\cite{du02} and \cite{vitkalov06} seem
quite different from each other.

The goal of our work is to show that the two seemingly different
phenomena are essentially two manifestations of the electron kinetics
described by a standard Boltzmann equation for a weakly disordered
2DEG in the presence of electric and magnetic fields. We demonstrate
that the low-current nonlinearity~\cite{vitkalov06} is the consequence
of the variation of the occupation factors of electron states: the
non-equilibrium population of states renders the transitions normally
contributing to the dissipative current ineffective. At high currents,
the effect of the electric and magnetic fields on the electron motion
becomes important. The oscillations in the $I$-$V$ characteristic are
associated with the geometric resonance in the electron transitions.

We evaluate the dissipative component of the electric current density
in a 2DEG placed in a perpendicular magnetic field $B$ as a function
of electric field characterized by the dimensionless parameter
$\zeta$,
\be \zeta=\pi \frac{2 eE\Rc}{\hbar\wc},\ \ \ \Rc=\frac{v_{\rm
    F}}{\wc},\ \ \ \wc=\frac{eB}{m_{\rm e}c},
\label{zeta}
\ee
proportional to the ratio of the work of electric field
associated with the displacement of the guiding center of
a cyclotron trajectory by $2\Rc$ to the Landau quantization energy
$\hbar\wc$. The displacement occurs due to the electron scattering off
an impurity, and does not exceed $2\Rc$ in a single scattering act.
This geometrical constraint leads to the oscillations of the current
with $\zeta$; each oscillation corresponds to an increase by
$\hbar\wc$ of the maximal energy acquired by an electron from the
electric field in a single scattering event. The maximal
displacement of the guiding center is reached for the scattering angle $\pi$
(backscattering), thus the amplitude of oscillations is proportional to
the corresponding scattering rate $1/\tau(\pi)$.

The ``preferred'' values of the energy absorbed by an electron from
the electric field are multiples of $\hbar\wc$ because of the
oscillations of the electron DOS associated with the
Landau quantization. It is interesting to note however, that even a
strong electric field does not result in developing a substantial
modulation in the electron energy distribution with the period
$\hbar\wc$. The reason for that is the dual role electric field
plays. On one hand, it promotes the build-up of electron distribution
at the energies corresponding to the maxima of the DOS.
On the other hand, it increases the electron diffusion in energy
space, the corresponding coefficient of the spectral diffusion being
proportional to the Joule losses. The latter effect wins over the
former one, and the electron distribution in energy gets smoother at
higher fields. As the result, the oscillatory part of the $I$-$V$
characteristic reflects the modulation of the electron transition
rates with the field, rather the modifications in the electron
distribution function; the amplitude of oscillations provides
information about the backscattering rate $1/\tau(\pi)$,
hardly accessible in other experiments.

In this work we express the dissipative current in terms  of the
inelastic relaxation rate $1/\tauin$ and harmonics $1/\tau_n$
of the elastic electron scattering rate $1/\tau(\theta)$ on angle $\theta$:
\be
\frac{1}{\tau(\theta)}=
\sum_{n=-\infty}^{+\infty}
\frac{e^{in\theta}}{\tau_{n}},\quad
\tau_n=\tau_{-n}.
\label{tau}
\ee
%with $\tau_{n}=\tau_{-n}$.
Typically, the ``quantum scattering time'' $\tauq$ is short,
$\tau_0\ll \tau_{\rm in}$. However, the transport relaxation time,
defined as $1/\tautr=1/\tauq-1/\tau_1$, may be in an arbitrary
relation with $\tau_{\rm in}$. We show that the measurements of the
dissipative current as a function of $\zeta$ at small ($\zeta \lesssim
\sqrt{\tauq/\tauin}$) and large ($\gtrsim 1$) values of $\zeta$
reveal the rates of inelastic relaxation and of the back-scattering
off disorder, respectively.

In the following, we consider the limit of high temperatures $T\gtrsim
\hbar\wc/2\pi^2$, when the non-linear resistivity is observed, but the
SdH oscillations are already suppressed~\cite{du02}. We also limit our
analysis to the case of ``classically strong'' magnetic fields, {\it
  i.e.}, we assume that $\wc\tauq\lesssim 1$ while $\wc\tautr\gg
1$. The former condition allows us to keep only the first harmonic in
the oscillations of the DOS
\be
\begin{split}
&\nu(\vare)= \nu_0\left(1-2\lambda\cos\frac{2\pi\vare}{\hbar\wc}
\right),\quad \lambda=  e^{-\pi/\wc\tauq}. \label{dos}
\end{split}
\ee
The condition $\tautr/\tauq\gg 1$ is routinely met in semiconductor
heterostructures, and the domain of magnetic fields $1/\tautr\ll \wc\ll
1/\tauq$ is quite wide.

\begin{figure}
\epsfxsize=0.25\textwidth
%\vspace*{0.3\textwidth}
\centerline{\epsfbox{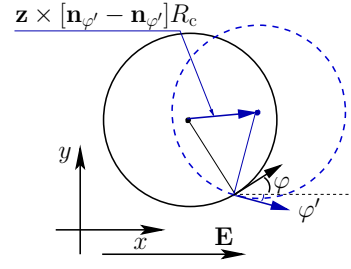}}
\caption{Electron scattering off impurity changes the momentum direction from
$\bm{n}_\varphi$ to $\bm{n}_{\varphi'}$ and the position of the guiding center
shifts by $\Rc\bm{z}\times[\bm{n}_{\varphi'}-\bm{n}_\varphi]$.}
\label{fig:explanation}
\end{figure}

The dissipative part of the electric current
\be
j_{\rm d}=2e\vF\int d\vare \nu(\vare)
\int \cos\varphi
f(\vare,\varphi)\frac{d\varphi}{2\pi}
\label{jdc}
\ee
is determined by the stationary electron distribution function
$f(\vare,\varphi)$, which is the solution of the following
kinetic equation
\be
-\wc\partial_\varphi f(\vare,\varphi)=
{\rm St}_{\varphi}\{f(\vare,\varphi)\}+
{\rm St}_{\rm in}\{f(\vare,\varphi)\}.
\label{kineq}
\ee
Here $\varphi$ is the angle the electron momentum makes with the
direction of the electric field. The first term in the right hand side
of \req{kineq} is the collision integral for electron scattering off
disorder:
\be
\begin{split}
{\rm St}_{\varphi}  \{f\}=
\int\frac{
\nu(\vare+ W_{\varphi\varphi'})}{\nu_0}
\frac{f(\vare+ W_{\varphi\varphi'},\varphi')
-
f(\vare,{\varphi})}{\tau(\varphi-\varphi')}\frac{d\varphi'}{2\pi}.
\label{Stel}
\end{split}
\ee
Here $W_{\varphi\varphi'}=eE \Rc [\sin{\varphi'}-\sin\varphi]$
is the work of the electric field in the course of the shift
$\Rc \bm{z}\times[\bm{n}_{\varphi'}-\bm{n}_\varphi]$ of the
guiding center of the cyclotron trajectory, see Fig.~\ref{fig:explanation};
unit vector $\bm{z}$ is perpendicular to the 2DEG plane, and
$\bm{n}_\varphi=\{\cos\varphi,\ \sin\varphi\}$ is directed along the electron momentum.
The rate of such scattering events is given by $1/\tau(\varphi'-\varphi)$ and
characterized by its harmonics $1/\tau_n$, see \req{tau}.
The imbalance between scattering ``in'' and ``out'' terms is determined
by the difference of the corresponding distribution functions
$f(\vare+W_{\varphi\varphi'},\varphi')$ and
$f(\vare,\varphi)$.

The inelastic relaxation at sufficiently low temperatures is
dominated by the electron-electron interaction, represented by
the inelastic collision integral
\begin{eqnarray}
&&{\rm St}_{\rm in}\left\{ f (\vare) \right\}  =
\int d\vare'\int dE   M(E, \vare, \vare')
\label{Stin}
\\
&&\times \left[
\tilde  f (\vare)f(\vare_+)\tilde f (\vare')
f(\vare'_-)
-f(\vare)\tilde f (\vare_+)f(\vare') \tilde f (\vare'_-)
\right].
\nonumber
\end{eqnarray}
Here $\tilde f(\vare) \equiv 1-f(\vare)$, $\vare_{+}= \vare + E$, $\vare_{-}'=
\vare' - E$ and $M(E,\,\vare,\,\vare')$ describes the
dependence of the matrix element of the screened Coulomb interaction on the
transferred energy $E$ and the electron energies $\vare$ and
$\vare'$.

To the first order in  $1/\wc\tautr\ll 1$ we look for a solution
of the kinetic equation~\req{kineq} in the form:
\begin{eqnarray}
&&f(\vare,\varphi)=
\left[1-\frac{eE\cos\varphi \Rc^2}{\vF\tautr}\partial_\vare\right]
f_{T}(\vare)
\label{fgen}
\\
&&
+\lambda \left\{I_1\sin\frac{2\pi \vare}{\hbar\wc}
+\left[A_1
\cos\frac{2\pi \vare}{\hbar\wc}+\lambda A_2\right]\cos\varphi
\right\} \partial_\vare f_{T}(\vare).
\nonumber
\end{eqnarray}
The first term in \req{fgen} is the result for $\lambda=0$, when the
effect of quantization by magnetic field is neglected. The second term
in \req{fgen} is proportional to $\lambda$ and oscillates with
$\vare/\wc$. Here we assume  that $T\gg \wc$, so that we can
separate fast oscillatory dependence on energy of $f(\vare,\varphi)$
with period $\hbar\wc$ and smooth energy dependence of
$f_T(\vare)$ on the scale
of temperature $T$ of the 2DEG. We assume that $\hbar\wc \zeta\ll
T$ and that $f_{T}(\vare)$ is close to the Fermi distribution
function.

The oscillations amplitude $I_1(\vare)$
of isotropic in momentum space component of the distribution function
\req{fgen} is determined by
\begin{subequations}
\label{kineq:amp}
\be
\frac{I_1(\vare)}{ \tau_{\rm in}(\vare)}\sin\frac{2\pi\vare}{\hbar\wc}=
\left\langle{\cal K}^{(0)}_{\varphi\varphi'}(\vare)  I_1
+
{\cal K}^{(1)}_{\varphi\varphi'}(\vare)
\right\rangle
.
\label{kineq:iso}
\ee
For the amplitudes $A_{1,2}$ of the anisotropic component in \req{fgen}
we have
\be
\frac{\wc}{2}A_1\cos\frac{2\pi\vare}{\hbar\wc}=
\left\langle \sin\varphi\left[{\cal K}^{(0)}_{\varphi\varphi'}(\vare)
I_1 +
{\cal K}^{(1)}_{\varphi\varphi'}(\vare)  \right]
\right\rangle,
\label{kineq:j}
\ee
\be
\frac{\wc}{4}A_2=
-\left\langle \sin\varphi\cos\frac{2\pi(\vare+W_{\varphi\varphi'})}
{\hbar\wc}{\cal K}^{(0)}_{\varphi\varphi'}(\vare)
\right\rangle I_1.
\label{kineq:j2}
\ee
\end{subequations}
Here $\langle\dots \rangle$ stands for averaging over
angle variables $\varphi$ and $\varphi'$.
The kernels ${\cal K}^{(0,1)}_{\varphi\varphi'}(\vare)$ are given by
\be
\begin{split}
{\cal K}^{(0)}_{\varphi\varphi'}(\vare) & =
 \frac{\sin[2\pi(\vare+W_{\varphi\varphi'})/\hbar\wc]
-\sin[2\pi\vare/\hbar\wc]
}{\tau(\varphi-\varphi')},
\\
{\cal K}^{(1)}_{\varphi\varphi'}(\vare) & =-2
 \cos\frac{2\pi(\vare+W_{\varphi\varphi'})}{\hbar\wc}
\frac{W_{\varphi\varphi'}} {\tau(\varphi-\varphi')}.
\end{split}
\ee
In the left hand side of \req{kineq:iso} we used the linearized
form of the inelastic collision integral \req{Stin}. The
inelastic relaxation rate $1/\tauin(\vare)$ of the oscillating
component of the distribution function \req{fgen} was
calculated in~\cite{DVAMP}:
\be
\frac{1}{\tauin(\vare)}=\frac{\pi^2T^2+\vare^2}{4\pi E_{\rm F}}
\ln\frac{\kappa v_{\rm F}}{{\rm max}\{T,\sqrt{\wc^{3}\tautr}\}},
\ee
where $E_{\rm F}$ is the Fermi energy and $\kappa=4\pi e^2\nu_0$.
Solving \req{kineq:iso} with respect to $I_1$, we obtain
\begin{subequations}
\label{fresult}
\be
I_1(\vare)  =   \frac{-2 eE\Rc[d\gamma(\zeta)/d\zeta]}
{\tauin^{-1}(\vare)+\tauq^{-1}-\gamma(\zeta)},\ \
\gamma(\zeta)=\sum_n\frac{J_n^2(\zeta)}{\tau_n}.
\label{fis}
\ee
Next we substitute $I_1(\vare)$ into \reqs{kineq:j} and
\rref{kineq:j2} and find
\be
\begin{split}
A_1(\vare) & = -(2eE\Rc/\wc\tautr)
\left[\Gamma_{1}(\zeta,\tauin(\vare))+\Gamma_{2}(\zeta)
\right],\\
A_2(\vare) & =(2eE\Rc/\wc\tautr) \Gamma_{1}(\zeta,\tauin(\vare))
\label{fan}.
\end{split}
\ee
\end{subequations}
Here we introduced the following notations:
\be
\frac{\Gamma_{1}(\zeta,\tauin)}{\tautr} =-\frac{[d\gamma(\zeta)/d\zeta]^2}
{\tauin^{-1}+\tauq^{-1}-\gamma(\zeta)},\ \
\frac{\Gamma_{2}(\zeta)}{\tautr}  = - \frac{d^2\gamma(\zeta)}{d\zeta^2},
\label{G12}
\ee
and $J_n(\zeta)$ are the Bessel functions.

The isotropic in momentum and oscillatory in energy
component of the distribution function $f(\vare,\varphi)$
results in the $\Gamma_1(\zeta,\tauin)$ contribution
to the amplitudes $A_{1,2}$ of anisotropic part
of  $f(\vare,\varphi)$~\cite{DVAMP}. The second contribution,
containing $\Gamma_2(\zeta)$, is coming from the second term in
the r.h.s. of \req{kineq:j}. This contribution was studied in
Refs.~\cite{ryzhii,durst03,VA03} and originates directly from the
effects of electric fields on the
collision integral for scattering off disorder.

Substituting the distribution function \req{fgen} with oscillating components
\req{fresult} into \req{jdc} and integrating over energy,
we obtain the dissipative current:
\be
j_{\rm d}=\sigma_{\rm D}%\frac{e^2 \Rc^2\nu_0}{2\tautr}
E[1+2\lambda^2 F(\zeta)],\ \
F(\zeta)=2\Gamma_{1}(\zeta,\tauint)+
\Gamma_{2}(\zeta),% \sigma_{\rm D}=\frac{e^2 \Rc^2\nu_0}{2\tautr}
\label{current}
\ee
where $\sigma_{\rm D}=e^2\Rc^2\nu_0/2\tautr$ is the Drude
conductivity at large Hall angle, $\wc\tautr\gg 1$.
To perform the integration and
simplify further analysis we replaced the inelastic relaxation rate
$1/\tauin(\vare)$ by a parameter $1/\tauint$, which can be chosen as
$1/\tauint\sim 1/\tauin(\vare=T)$.

Equation~\rref{current} for the dissipative current together with
\reqs{fis} and \rref{G12} for functions $\gamma(\zeta)$, $\Gamma_1(\zeta,\tauin)$
and $\Gamma_2(\zeta)$ constitute
the central result of the paper. Below we discuss the properties
of functions $\Gamma_1(\zeta,\tauint)$ and $\Gamma_2(\zeta)$, which
determine the non-linear response of the dissipative current $j_{\rm
d}$, \req{current}. Then we consider a specific model for
$1/\tau_n$ in \req{tau} and analyze the
non-linear behavior of the current within the model.

For weak electric fields, $\zeta\ll 1$, we obtain the following
expressions for functions $\Gamma_{1}(\zeta,\tauint)$ and
$\Gamma_{2}(\zeta)$ expanding the Bessel functions to lowest order in
$\zeta^2$:
\begin{subequations}
\label{smallz}
\begin{eqnarray}
\Gamma_{1}(\zeta,\tauint)
& = & -\frac{(\tauint/\tautr)\zeta^2}{1+(\tauint/2\tautr)\zeta^2},
\label{smallz:k}
\\
\Gamma_{2}(\zeta) & = & 1-
\frac{3}{8}\tautr\zeta^2\left[\frac{3}{\tau_0}-\frac{4}{\tau_1}+\frac{1}{\tau_2}
\right].
\label{smallz:d}
\end{eqnarray}
\end{subequations}
The $\zeta^2$ term in the denominator of \req{smallz:k} is
legitimate in the limit $\tauin\gg\tautr$, which may take place at
sufficiently low electron temperatures. We also note that
\req{smallz:k} coincides with the result of Ref.~\cite{DVAMP}
in the absence of microwave fields.

In the strong-field limit, $\zeta\gg 1$, we find
\be
\frac{\Gamma_{1}(\zeta,\tauint)}{\tautr}
\propto -\frac{\tau_0}{\tau^2(\pi)}
\frac{\cos^2\zeta}{\zeta^2}, \ \
\frac{\Gamma_{2}(\zeta)}{\tautr}
\propto \frac{1}{\tau(\pi)}\frac{\sin
2\zeta}{\zeta},
\label{asymptotes}
\ee
where $1/\tau(\pi)=\sum_n e^{i\pi n}/\tau_n$ is the
back-scattering rate off disorder. The $\Gamma_2$-contribution,
arising from the effect of electric field on the collision integral,
is larger than the $\Gamma_1$-contribution, which arises from the
stationary out-of-equilibrium component of the distribution function.
The latter contribution not only decays faster with the increase of
$\zeta$ than the former one, but also contains an additional
parameter $\tauq/\tau(\pi)$ which is small for smooth disorder. The
amplitude of oscillations of current \req{current} decays
proportionally to $1/\zeta$ at $\zeta\gg 1$, but the oscillations in
differential conductivity
$$
\sigma=\partial j_{\rm d}/\partial E \propto \partial [\zeta\Gamma_2(\zeta)]/\partial
\zeta\propto \cos 2\zeta
$$
do not vanish; its maxima and minima are situated at $\zeta=\pi k/2$ with integer
$k$. At smaller values of $\zeta$, term $\Gamma_1(\zeta)$ also
contributes to $\sigma$ and results in dependence of the oscillations amplitude
on $\zeta$ as well as in some shift of maxima and minima from $\zeta=\pi
k/2$.

To discuss the properties of the non-linear current \req{current} in a
broad range of electric fields, we consider a specific model for the
harmonics $1/\tau_n$ of the scattering rate off disordered potential
due to charged impurities inside or in the proximity of 2DEG:
\be
\frac{1}{\tau_{n}}=
\frac{1}{\taul}\frac{1}{1+\chi n^2}+
\frac{\delta_{n,0}}{\taus},\quad \chi\ll 1.
\label{taun}
\ee
Here $1/\taus$ is the scattering rate off impurities inside the 2DEG,
which produce sharp ($\delta$-correlated) potential for
electrons. Charged impurities in the proximity of
2DEG produce a smooth potential  resulting in
electron scattering on small angle $\theta \sim \sqrt{\chi}\ll 1$,
where $\chi$ can be estimated as $\chi\sim (\lambda_{\rm F}/\xi)^2$,
with $\lambda_{\rm F}$ and $\xi$ being the Fermi wavelength
and the correlation length of the disorder potential, respectively.
The two restrictions for validity of \req{current}
on the strength of magnetic field
($\wc\tautr\gg 1$ and $\wc\tauq\lesssim 1$)
can be satisfied simultaneously for $\chi\ll 1$.

\begin{figure}
\epsfxsize=0.35\textwidth
%\vspace*{0.3\textwidth}
\centerline{\epsfbox{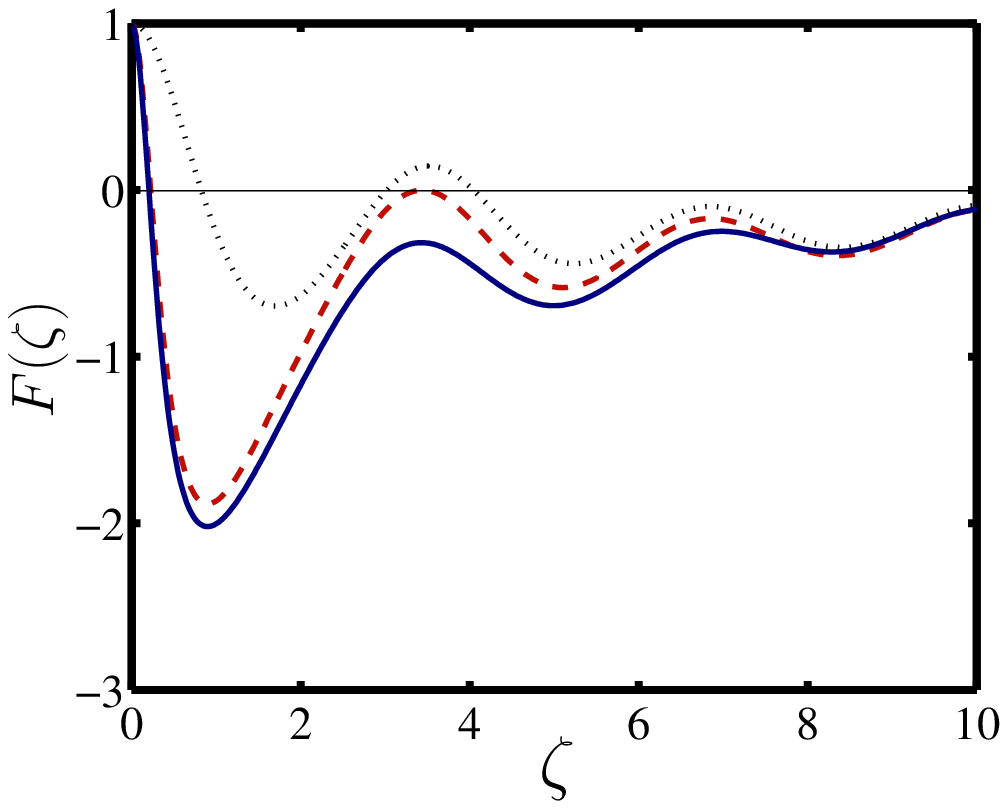}}
\caption{Function $F(\zeta)$ for disorder
  described by \req{taun} with $\taul=\taus/30$ for several values of
  $\tauint$ and $\chi$: $\tauint=10\taus$, $\chi=0.02$ (solid);
  $\tauint=10\taus$, $\chi=0.01$ (dashed); $\tauint=0.5\taus$, $\chi=0.01$
  (dotted).  Note that $F(\zeta)$ has a peak at small $\zeta$
  determined by the inelastic relaxation time $\tauint$ and oscillates
  at large $\zeta$.  At intermediate
  $\zeta\sim 1$ the positions of maxima and minima of $F(\zeta)$ depend
  on the relation between various parameters of the model.}
\label{fig3}
\end{figure}

For the disorder characterized by harmonics of the scattering rate
\req{taun}, we obtain $ \gamma(\zeta)=J_0^2(\zeta)/\taus+
1/(\taul\sqrt{1+\chi\zeta^2})$. Here we omitted the term, which
arises from the back-scattering off smooth disorder and yields
exponentially small ($\exp(-\pi/\sqrt{\chi})$) contribution to
$1/\tau(\pi)$, cf.  \req{asymptotes}. Substituting this expression for
$\gamma(\zeta)$ in~\req{G12}, we can evaluate the
current~\req{current} at arbitrary $\zeta$. At $\zeta\ll
1/\sqrt{\chi}$ we have
\be \frac{F(\zeta)}{\tautr}
=\frac{1}{\tautrl}
-\frac{2[2J_0(\zeta)J_1(\zeta)/\taus+\zeta/\tautrl]^2}
{\tauint^{-1}+\taus^{-1}\![1-J_0^2(\zeta)]+\tautrl^{-1}\zeta^2/2}-
\frac{[J_0^2(\zeta)]''}{\taus},
\label{broad}
\ee
where $1/\tautrl=\chi/\taul$ is the smooth disorder contribution to
the transport scattering rate.
Equation~\rref{broad} covers both the regime of relatively weak fields,
where inelastic scattering is important, and the regime of strong
fields, exhibiting prominent oscillations. Function $F(\zeta)$ has a
sharp feature at $\zeta\sim \sqrt{\taus/\tauin}$. At these fields, the
spectral diffusion of electrons caused by electric field becomes
comparable with the inelastic relaxation. At stronger fields, $\zeta\gtrsim
\sqrt{\tautrl/\tauin}$, the two kinds of disorder, smooth and
``sharp'', yield two separate contributions to $F(\zeta)$,
\be
%\begin{split}
 \frac{F(\zeta)}{\tautr} =
 \frac{1-3\chi\zeta^2-2(1+\sqrt{1+\chi\zeta^2}) }{\tautrl(1+\chi\zeta^2)^{5/2}}
 -\frac{[J_0^2(\zeta)]''}{\taus}.
%\end{split}
\ee
Here only the sharp component of disorder
contributes to the oscillatory behavior of $F(\zeta)$.

We calculated the dissipative
component of electric current in response to the applied dc electric
field of arbitrary strength within self-consistent Born approximation
\cite{Ando}. We show that the non-linear component of the current
consists of two contributions. One contribution arises due to the
formation of the out-of-equilibrium component of the distribution
function, oscillating as a function of energy. The second contribution
is the result of modification by electric field of electron scattering
amplitudes off the disorder potential.  We showed that the former
contribution is important at relatively weak fields, while the
latter one dominates in the high-field domain.  There, the non-linear
contribution to the current oscillates as a function of the applied
electric field.  The amplitude of oscillations of the differential
conductivity does not decrease with the increase of electric field
(and at fixed magnetic field).  It may be necessary to take into
account the effect of heating on the quantum scattering time and thus
on the Dingle factor $\lambda$ in order to explain the suppression of
oscillations observed in~\cite{zudov06}.

Finally, we considered the limit $\wc\tauq\lesssim 1$, and therefore
assumed that the oscillations in the DOS are described by one harmonic
with period in energy $\hbar\wc$, \req{dos}. In stronger fields, the
DOS remains periodic in energy with the same period $\hbar\wc$, but
contains higher harmonics~\cite{Ando,VA03}. These higher harmonics in
the DOS result in a more complicated form of the oscillatory part of
the non-linear resistivity.

The authors are thankful to S. Vitkalov
and M. Zudov for discussions of experiments.  Part of this work
was performed during the visit to the Aspen Center for Physics. The
work was supported by  the W. M. Keck Foundation and NSF grants
DMR 02-37296, DMR 04-39026 and DMR-0408638.

%%%%%%
%%%%%%\bibliographystyle{prsty}
%%%%%%\bibliographystyle{prsty}
%%%\bibliography{NL2DEG}

\end{document}